\pdfoutput=1

\documentclass[11pt,twoside,a4paper,cmspaper,final,collab]{cms-tdr}
\def\svnVersion{494420P}\def\cmsCernNoTag{CERN-EP-2018-308}\def\cmsCernDate{\today}\def\cmsMessage{Published in the European Physical Journal C as \href{http://dx.doi.org/10.1140/epjc/s10052-019-6861-x}{\doi{10.1140/epjc/s10052-019-6861-x.}}}

\begin{document}\cmsNoteHeader{FSQ-15-006}

\hyphenation{had-ron-i-za-tion}
\hyphenation{cal-or-i-me-ter}
\hyphenation{de-vices}
\RCS$HeadURL: svn+ssh://svn.cern.ch/reps/tdr2/papers/FSQ-15-006/trunk/FSQ-15-006.tex $
\RCS$Id: FSQ-15-006.tex 494413 2019-05-09 09:25:33Z salim $

\newcommand {\Cas}{CASTOR\xspace}
\newcommand {\EPOS}{\textsc{epos-lhc}\xspace}
\newcommand {\SIBYLL}{\textsc{sibyll}\,2.1\xspace}
\newcommand {\QGSJET}{\textsc{qgsjetII{.}04}\xspace}
\newcommand {\QGSJETONE}{\textsc{qgsjet01}\xspace}
\newcommand {\monash}{\textsc{Monash 2013}\xspace}
\newcommand {\INEL}{inclusive inelastic\xspace}
\newcommand {\sqrts}{\sqrt{s}\xspace}
\newcommand {\NSD}{non-single-diffractive-enhanced\xspace}
\newcommand {\SD}{single-diffractive-enhanced\xspace}
\newcommand {\PU}{pileup\xspace}
\newcommand {\errscalehf}{10\xspace}
\newcommand {\errscalecas}{17\xspace}
\newcommand {\pois}[2]{\ensuremath{\operatorname{Pois} \left({#1}; {#2}\right)}}
\newcommand {\dhs}[1]{\ensuremath{\rd#1}}
\newcommand {\DHS}[1]{\ensuremath{\Delta#1}}
\newcommand {\dhe}[2]{\ensuremath\frac{\dhs{#1}}{\dhs{#2}}}
\newcommand {\dhi}[2]{\ensuremath\dhs{#1}/\dhs{#2}}
\newcommand {\yb}{\ensuremath y_{\mathrm{beam}}}
\newcommand {\Et}{\ensuremath E_{\mathrm{T}}}
\newcommand {\etamin}{\ensuremath3.15}
\newcommand {\etamax}{\ensuremath6.6}
\newcommand {\etamincas}{\ensuremath5.2}
\newcommand {\etamaxhf}{\ensuremath5.2}

\cmsNoteHeader{FSQ-15-006}
\title{Measurement of the energy
       density as a function of pseudorapidity  in proton-proton collisions at $\sqrts=13\TeV$}

\date{\today}

\abstract{A measurement of the energy density in proton-proton
    collisions at a centre-of-mass energy of $\sqrts=13$\TeV is
  presented.  The data have been recorded with the CMS experiment
  at the LHC during low luminosity
  operations in 2015. The energy density is  studied as a function
  of pseudorapidity in the ranges
  $-6.6<\eta<-5.2$ and $3.15<\abs{\eta}<5.20$. The results are compared with
  the predictions of several models.
  All the  models considered suggest a different shape of the
  pseudorapidity dependence compared to that
  observed in the data.
  A comparison with LHC proton-proton
  collision data at $\sqrts=0.9$ and $7\TeV$
  confirms the compatibility of the data with the hypothesis of
  limiting fragmentation.  }

\hypersetup{
  pdfauthor={CMS Collaboration},
  pdftitle={Measurement of the energy
       density as a function of pseudorapidity  in proton-proton collisions  at sqrt(s) = 13 TeV},
  pdfsubject={CMS},
  pdfkeywords={CMS, physics, small-x QCD, underlying event, multi-parton interactions, HF, CASTOR}}

\maketitle

\section{Introduction}

In the framework of quantum chromodynamics (QCD), inelastic
proton-proton collisions are described by a combination of hard and soft exchanges between the constituents of the protons. Hard collisions between
one or multiple pairs of partons are complemented by soft parton
scattering from Multiple Parton Interactions
(MPI)~\cite{Sjostrand:1986ep, Sjostrand:1987su, Borozan:2002fk,
  Sjostrand:2004pf}, parton shower effects including initial- and final-state radiation, which, along with
projectile fragmentation, constitute the \emph{underlying event} (cf.~Ref.~\cite{Khachatryan:2010pv}).
At the CERN LHC these effects can be studied at the
highest possible centre-of-mass energies covering a very large angular
phase space. The measurement of the average energy per proton-proton collision in different
pseudorapidity ($\eta$) regions probes our general understanding of
QCD multiparticle production. Moreover, because of the extended calorimetric instrumentation of the
CMS experiment beyond $\abs{\eta}>3$, covering the full range from
$-\etamax$ to $+\etamaxhf$ in pseudorapidity, smaller scattering
angles may be accessed compared to other measurements.

In this paper, a measurement of the energy density in proton-proton
collisions at the centre-of-mass energy $\sqrts=13\TeV$ within the pseudorapidity ranges
 $-6.6<\eta<-5.2$ and $3.15<\abs{\eta}<5.20$ is presented.
This measurement extends the $\sqrts$ and pseudorapidity range covered by
previous results from the CMS~\cite{Chatrchyan:2011wm}, ATLAS~\cite{Aad:2012mfa}, and
LHCb~\cite{Aaij:2012pda} Collaborations.
The average energy density per collision is defined as
\begin{equation}
  \dhe{E}{\eta}=
  \frac{1}{N_\text{coll}} \sum_{i}{E_{i}} \frac{c(\eta)}{\Delta\eta},
  \label{eq:enfl}
\end{equation}
where $\sum_{i} E_{i}$ is the summed energy measurements of all calorimeter
towers $i$ within a bin of pseudorapidity having a width
$\Delta\eta$,
$c(\eta)$ is the $\eta$-dependent conversion factor from
the calorimeter measurements to a stable-particle level energy,
and $N_\text{coll}$ is the number of selected proton-proton collisions corrected for the
contributions from noise and simultaneous {\Pp\Pp} collisions occurring in
the same event (\PU). By \emph{event} we refer to the data of one single LHC bunch crossing.
To investigate various aspects of MPIs in high-energy proton-proton
collisions the measurement is performed for several different
categories of collision, each category defined by a specific event selection.

Moreover, the data collected at $\sqrts=13$\TeV are analysed together with data
collected at 0.9 and 7\TeV~\cite{Chatrchyan:2011wm}.  This is
interesting since projectile fragmentation can then be studied in the regions
close to the beam rapidity, $\yb=\text{acosh}(\sqrts/2m_\text{p})$, where
 $m_\text{p}$ is the mass of the projectile particle, \ie a proton in
the present case. At $\sqrts=13\TeV$, $\yb\approx9.5$, while at $\sqrts=0.9\TeV$ it is
just $\approx$6.8. Thus, the detectors of CMS, although located at fixed
$\eta$, cover a very wide range in $\eta'=\eta-\yb$ when data recorded at
different centre-of-mass energies are combined. The hypothesis of limiting fragmentation~\cite{Benecke:1969sh}
suggests that particle production reveals longitudinal scaling,
\ie the dependence of very forward particle production on the
centre-of-mass energy vanishes in the region $\eta'\approx0$~\cite{Ruan:2010ig}.
In this paper, the hypothesis
of limiting fragmentation is tested in collisions at $\sqrts$ from 0.9 to 13\,\TeV.

Measurements of the energy density at collider energies are an
important reference necessary for extrapolating to even  higher
centre-of-mass energies.
The results reported here provide valuable input for the tuning of Monte
Carlo models used to describe the highest energy hadronic interactions
needed for the interpretation of cosmic ray
measurements~\cite{Ulrich:2010rg,dEnterria201198}.

\section{The CMS detector}

At the heart of the CMS detector is a superconducting solenoid of 6\unit{m}
internal diameter, providing a strong magnetic field of
3.8\unit{T}. The data used for this paper were taken in
June 2015 during a period
without magnetic field.
Within the CMS magnet volume are an inner silicon pixel and strip
tracker that measure charged particles in the range $\abs{\eta}<2.5$, a
homogeneous lead tungstate crystal electromagnetic calorimeter, and a
brass and scintillator hadron calorimeter.
The corresponding endcap detectors instrument the
pseudorapidity range up to $\abs{\eta}\lesssim3$ with tracking and calorimetry.
Forward
Cherenkov calorimeters extend the coverage beyond $\abs{\eta}\gtrsim3$.
Muons are measured in gas-ionization detectors embedded in the
steel return yoke.

The hadron forward (HF) calorimeters cover the region
$2.9<\abs{\eta}<\etamaxhf$ and consist of 2$\times$432 readout towers, each
containing a long and a short quartz fiber embedded within a steel absorber
running parallel to the beam. The long fibers run the entire depth of the HF calorimeter
(165\unit{cm}, or approximately 10 interaction length), while the
short fibers start at a depth of 22\unit{cm} from the front of the
detector. The response of each tower is determined from the sum of signal in
the corresponding long and short fiber.
There are 13 rings of towers in $\abs{\eta}$, each
with a size of $\DHS{\eta}\simeq0.175$, except for the
lowest and highest $\abs{\eta}$ rings, which
have a size $\DHS{\eta}\simeq0.11$ and $\DHS{\eta}\simeq0.30$, respectively.
The azimuthal segmentation of all towers is $10^\circ$, except for the one
at highest $\abs{\eta}$, which has $\DHS{\varphi} = 20^\circ$.

The very forward angles on one side of CMS ($-\etamax<\eta<-\etamincas$) are covered
by the \Cas calorimeter.  It has 16 azimuthal towers, each built from 14
longitudinal modules. The 2 front modules form the electromagnetic section,
and the 12 rear modules form the hadronic section.
The calorimeter is made of stacks of tungsten and quartz plates, read out by
PMTs, in two half-cylindrical mechanical structures, and is
placed around the beam pipe at a
distance of $-14.4$\,m away from the nominal interaction point.  The
overall longitudinal depth of both \Cas and HF corresponds to 10 hadronic
interaction lengths. The \Cas calorimeter is only operated during
periods of low LHC luminosity ($\mathcal{L_\text{inst}}<10^{30}\,\mathrm{cm}^{-2}\,\mathrm{s}^{-1}$)
since it cannot distinguish the secondaries from simultaneous \PU collisions.

The present analysis is restricted to the range of pseudorapidity
covered by the HF and \Cas calorimeters, excluding  the two
lowest $\abs{\eta}$ segments of the HF calorimeters because they are partially
located in the shadow of the endcap calorimeters.
This corresponds to a combined pseudorapidity
range of $\etamin<\abs{\eta}<\etamaxhf$~and~$-\etamax<\eta<-\etamincas$.
The analysis is performed using a data sample corresponding to an integrated luminosity
of 0.06\nbinv recorded with an average
number of proton-proton interactions per bunch
crossing of about 0.05.

A more detailed description of the CMS
detector can be found in Ref.~\cite{Chatrchyan:2008aa}.

\section{Monte Carlo models}

In this paper, various Monte Carlo event generators are used to correct the data
from detector- to stable-particle level and to compare with
the experimental results.

{\tolerance=1800
The \PYTHIA{8}~\cite{Sjostrand:2014zea} generator is a general purpose
Monte Carlo package that builds most of its predictive power upon
hard-scattering
matrix elements calculated in
perturbative QCD and parton showering
according to the Dokshitzer--Gribov--Lipatov--Altarelli--Parisi
  (DGLAP) \cite{Gribov:1972ri, Gribov:1972rt, Lipatov:1974qm,
  Altarelli:1977zs, Dokshitzer:1977sg} equations. The string
fragmentation model~\cite{Andersson:1983ia} is used for hadronization. The free parameters of
the simulations can be adjusted to describe measurements at different
centre-of-mass energies, resulting in the production of different so-called \emph{tunes} of
the model~\cite{Khachatryan:2015pea}.
\par}

{\tolerance=800
In this analysis, \PYTHIA{8} (version 8.212) is used together with the
\textsc{cuetp8m1}~\cite{Khachatryan:2015pea},
\textsc{cuetp8s1}~\cite{Khachatryan:2015pea},
and \monash~\cite{Skands:2014pea}
tunes, as well as with the
\textsc{mbr} model~\cite{Ciesielski:2012mc} combined with the
\textsc{4c}~\cite{Corke:2010yf} and \textsc{cuetp8m1} tunes. In
the \textsc{cuetp8m1} and \textsc{cuetp8s1} tunes, which are based on
the \monash and \textsc{4c} tunes, the parameters are adjusted to
describe underlying event measurements
from the Fermilab Tevatron and the LHC.  The tunes
are constructed using different
parton distribution function sets (\textsc{NNPDF2{.}3LO}~\cite{Ball:2012cx})
and \textsc{CTEQ6L1}~\cite{Pumplin:2002vw}, respectively).
\par}

The \EPOS~\cite{Pierog:2013ria}
and \QGSJET~\cite{Ostapchenko:2010vb} generators are commonly used to describe
extensive air showers in the atmosphere initiated by cosmic ray
particles, where soft physics is of primary importance. A combination of
Gribov--Regge multiple scattering~\cite{Drescher:2001}, perturbative
QCD, and string fragmentation are the cornerstones of both models.
While \QGSJET includes a small number of fundamental parameters, the
phenomenology implemented in \EPOS offers more opportunities for
tuning. In \EPOS a hydrodynamic, or collective, component is included
in a parametrised form~\cite{Pierog:2013ria}.

The collisions simulated with the \textsc{monash} and \textsc{mbr}
tunes of \PYTHIA{8}, and the \EPOS and \QGSJET event
generators, have been processed with a detailed simulation of the full
CMS detector based on \GEANTfour~\cite{Agostinelli:2002hh} and
reconstructed using the same software sequence that is used for
recorded collision events. These four models are used to correct for
detector effects.

\begin{figure*}[tb]
 \centering
 \includegraphics[width=.495\textwidth]{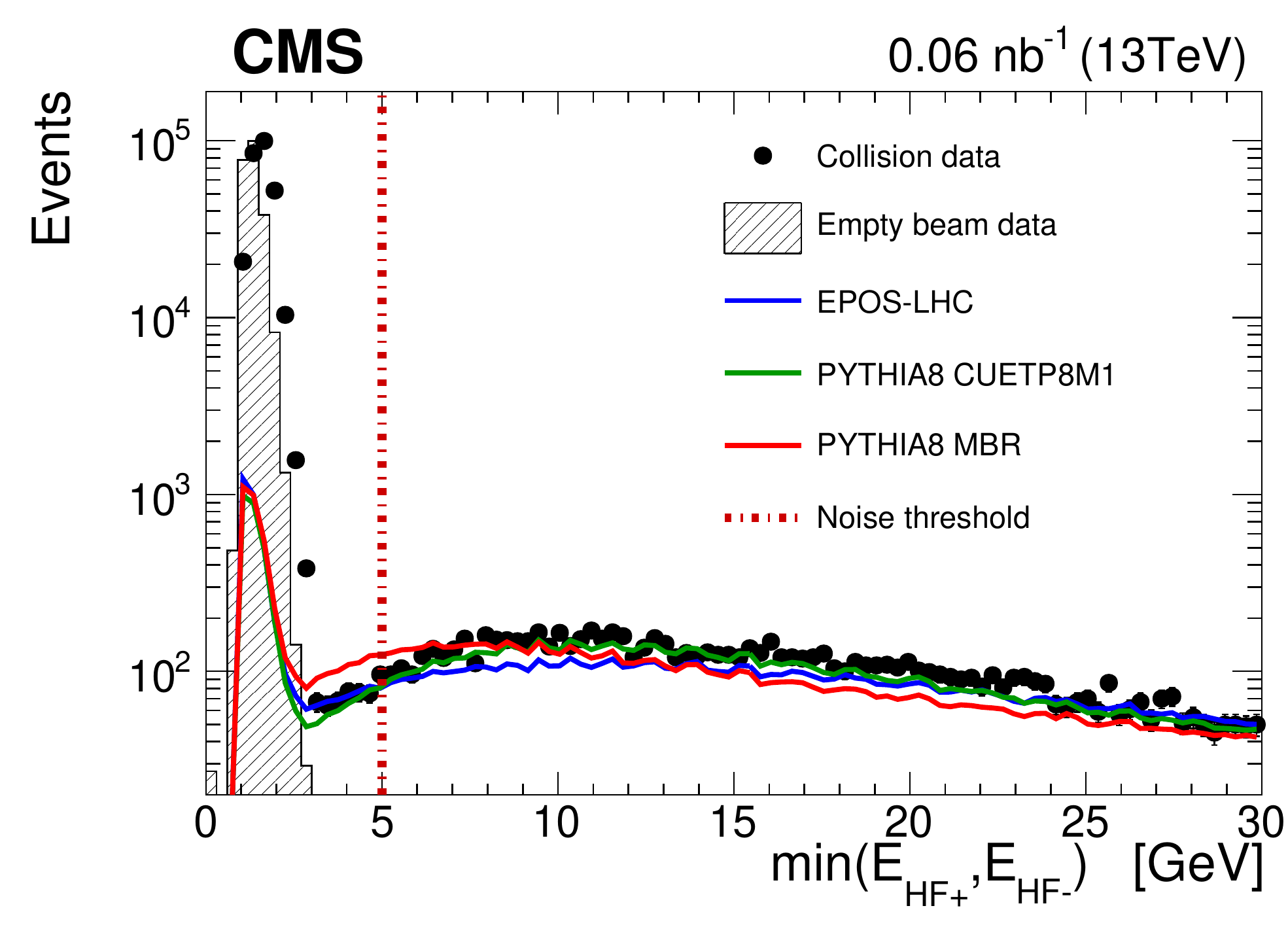}
 \includegraphics[width=.495\textwidth]{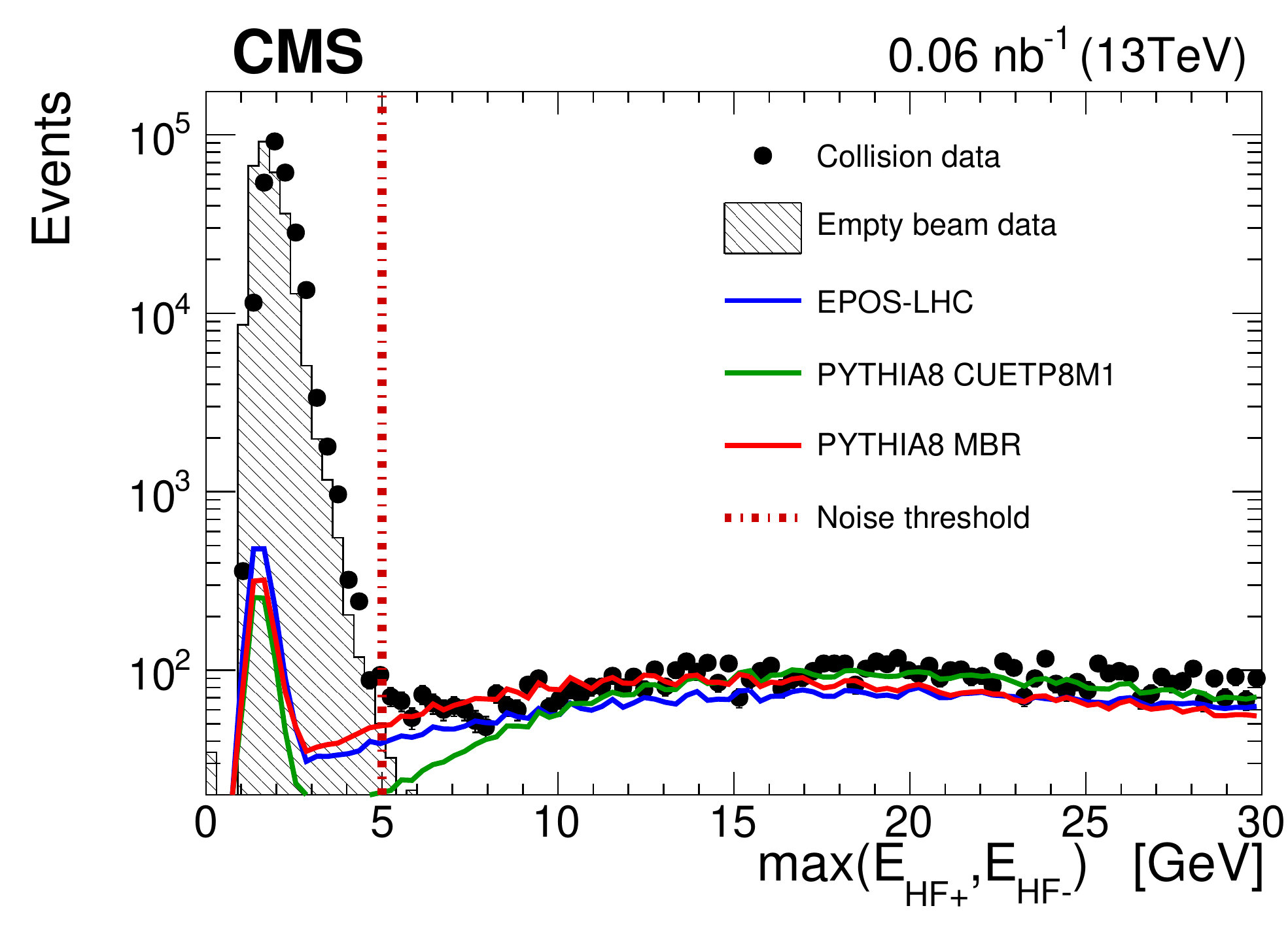}\\
 \caption{Distribution of the absolute
 number of events as a function of the highest energy tower, $E_\mathrm{HF+}$ and $E_\mathrm{HF-}$, in the
 HF$+$ and HF$-$ calorimeters.
 The left panel shows the smaller of the two HF calorimeter energies, min($E_\mathrm{HF-},E_\mathrm{HF+}$),
 whereas the right panel shows the higher of the two energies, max($E_\mathrm{HF-},E_\mathrm{HF+}$).
 The lines represent the simulations, while the markers represent the data.
 The measured detector noise distributions are shown as shaded areas. }
 \label{fig:noise}
\end{figure*}

\section{Event selection}

{\tolerance=1200
Events are selected online in an unbiased way by triggering the data
acquisition system with the Beam Pick-up-Timing for the
eXperiments (BPTX) devices~\cite{Khachatryan:2016bia}.
Three different categories of inelastic collisions are defined
offline: an {\textit{\INEL}} (INEL) selection to be as inclusive as possible, a
{\textit{\NSD}} (NSD-enhanced) selection, where single diffractive
dissociation contributions are suppressed, and a
{\textit{\SD}} (SD-enhanced) selection enriched in single diffractive
dissociation collisions. These selections are achieved by requiring an energy deposit
in the HF calorimeters above noise level either on at least one side (for the INEL
category) or on both sides (for the NSD-enhanced category), with respect to
the nominal interaction point of CMS. The SD-enhanced selection is
defined by requiring activity in one of the calorimeters on exactly one side,
with a veto condition being applied to the other side.
\par}

Energy deposition in the HF calorimeters is characterised by the
calorimeter tower with the highest energy in the negative (positive)
pseudorapidity region, $E_\mathrm{HF-}$ ($E_\mathrm{HF+}$), considering all
towers, except those belonging to the two rings closest to the endcap
(\ie at smallest $\abs{\eta}$). The energy thresholds for event
selection are determined from a study of events without beam and are
optimised to effectively reduce the contribution from detector noise,
while still allowing a high selection efficiency. In
Fig.~\ref{fig:noise}, the measured distributions for $E_\mathrm{HF-}$ and
$E_\mathrm{HF+}$ from collision data are shown together with the noise
distributions obtained from data without the presence of LHC beams.
This is achieved at the trigger level by requiring prescaled triggers where the two BPTX
 detectors are silent. In Fig.~\ref{fig:noise} simulated events
 are also shown. Events are selected for the INEL class if max($E_\mathrm{HF-}$,
 $E_\mathrm{HF+})>E_\text{threshold}$, and for the NSD-enhanced class if
min($E_\mathrm{HF-}$, $E_\mathrm{HF+})>E_\text{threshold}$. An energy
threshold of $E_\text{threshold}=5\GeV$ is found to be optimal to
suppress the noise contribution in both event classes for simulated
and measured events. For the NSD-enhanced category, the threshold could
in principle be lowered down to about $3\GeV$ without increasing
the noise contribution, but for consistency a
unified threshold of 5\GeV is used for all event classes.
The data were recorded at low luminosity with an interaction probability of about 5\%.
Most non-empty events contain a single proton-proton collision. A small fraction also
has two or more interactions. In contrast, the simulation was done without \PU, \ie, each
simulated event contains exactly one proton-proton collision.
The detector noise distribution as measured from empty-beam data
are also overlaid as shaded areas.

\begin{table*}[bt!]
  \centering
\topcaption{Summary of the event selections used for the different event categories in data at the detector level
  and in simulations at the stable-particle level. }
\begin{tabular}{p{2.9cm} p{5.6cm} p{5.2cm}}
\hline
Class & Detector level & Stable-particle level  \\
\hline
INEL & $E_\mathrm{HF+}>5\GeV$ or $E_\mathrm{HF-}>5\GeV$ &   $\xi>10^{-6}$\\\\
NSD-enhanced & $E_\mathrm{HF+}>5\GeV$ and $E_\mathrm{HF-}>5\GeV$
& at least one stable particle with $E>5\GeV$ in $-5.20<\eta<-3.15$ and $3.15<\eta<5.20$ \\\\
SD-enhanced & $E_\mathrm{HF+}>5\GeV$ and $E_\mathrm{HF-}<5\GeV$\newline
or\newline $E_\mathrm{HF+}<5\GeV$ and $E_\mathrm{HF-}>5\GeV$ & at least one
stable particle with $E>5\GeV$ in $3.15<\abs{\eta}<5.20$ on one side,
vetoing particles with $E>5\GeV$ on the other side\\\\
Limiting fragmentation study & $E_\mathrm{HF+}>4\GeV$ and $E_\mathrm{HF-}>4\GeV$ & one stable particle in $-4.4<\eta<-3.9$ and $3.9<\eta<4.4$ \\
\hline
 \end{tabular}
 \label{table:phasespace}
\end{table*}

\begin{table*}[bt!]
  \centering
  \topcaption{Selection factors and purities for various event selection
    categories. Only the first two parameters $f_\mathrm{EB}$ and $f_\mathrm{ZB}$ present actual measurements, from which the other quantities are derived
    as explained in the text.
    The probability $\epsilon$ to select a single collision
    is determined from simulations, and
    the value quoted here is the average value from all event
    generators, with a maximal model dependence of 2\%.
    The rightmost column quantifies the combined correction due to
    noise and \PU. All statistical uncertainties are negligible. }
  \label{tab:corr}
  \begin{tabular}{r  c c c c c  c}
        \hline
        & $f_\text{ZB}$ & $f_\text{EB}$ & $p$ & $\epsilon$ (MC) & $f_\text{PU}$ & $p\,f_\text{PU}$ \\
        \hline
        INEL & 0.0490 & 0.0005 & 0.9902 & 0.9051 & 1.0250  & 1.0149 \\
        HF+ & 0.0442 & 0.0003 & 0.9935 & 0.8224 & 1.0227  &  1.0161 \\
        HF$-$ & 0.0439 & 0.0002 & 0.9956 & 0.8232 & 1.0228 & 1.0183 \\\\
        NSD-enhanced & \NA & \NA & \NA & \NA & \NA & 1.0044 \\
        SD-enhanced & \NA & \NA & \NA & \NA & \NA & 0.9804 \\
        \hline
  \end{tabular}
\end{table*}

In simulated collisions particle four-momenta are used to build
sums of energies. At the stable-particle level (\ie for particles with proper decay length
c$\tau>1\cm$), simulated collisions are selected to be in the inclusive inelastic
category if $\xi=\text{max}(\xi_{\text{X}},\xi_{\text{Y}})>10^{-6}$, where
\begin{equation}
\xi_\mathrm{X}=\frac{M^2_{\text{X}}}{s},\qquad\qquad\xi_\mathrm{Y}=\frac{M^2_{\text{Y}}}{s},
\end{equation}
and $M_\mathrm{X}$ and $M_\mathrm{Y}$ are the invariant masses of the
particle systems on the negative and positive side of the largest rapidity
gap in the collision, respectively. This particular criterion for
stable-particle level is identical within a few percent with the
INEL detector level selection~\cite{Sirunyan:2018nqx}.

The NSD-enhanced collisions are selected at the stable-particle level with a
requirement of at least one stable particle
(either charged or neutral) within the
pseudorapidity acceptance of the HF calorimeters $\etamin<\abs{\eta}<\etamaxhf$ on both sides of the interaction point.

The SD-enhanced collision at the
stable-particle level are defined by the presence of
at least one stable particle with energy $E>5\GeV$ within the
pseudorapidity range $\etamin<\abs{\eta}<\etamaxhf$ on one side, whereas the
other side must be devoid of particles with energy $E>5\GeV$.

{\tolerance=5500
The phase space definitions for the NSD-enhanced, INEL and SD-enhanced
categories at the detector and
stable-particle level
 are summarised in Table~\ref{table:phasespace}.
 The last row of the table indicates  the
event selection needed for the limiting fragmentation study. This is
chosen to be identical to that used in previously published
data~\cite{Chatrchyan:2011wm} to allow a direct comparison of the results.
}

The energy density is measured with the HF and \Cas calorimeters by
summing up all the energy deposits in the calorimeter towers above noise
threshold. The value of the threshold was determined by measuring the detector noise and beam backgrounds using
empty-beam triggers (see Fig.~\ref{fig:noise} for HF results) and is
chosen to be $5\GeV$ in HF and $2.5\GeV$ in \Cas.
 The energy density measurement is performed as a function of $\abs{\eta}$. In the range
$\etamin<\abs{\eta}<\etamaxhf$ the corresponding measurements at positive
and negative pseudorapidities in HF are averaged, while for
$-6.6<\eta<-\etamincas$ the energy in \Cas is used. For the
SD-enhanced measurement only the side on which the HF calorimeter is
above noise level (thus, opposite to the forward rapidity gap) is used for the
measurement.

\section{Data analysis}

{\tolerance=800
The measurement of the energy density according to Eq.~(\ref{eq:enfl})
requires the determination of the number of selected collisions $N_\text{coll}$
and the energy sum, $\sum_i E_i$.
\par}

\begin{table*}[tb!]
  \centering
  \topcaption{The uncertainties in the
              energy density measurement for the three event selection categories.
              The results depend slightly on the pseudorapidity.
              }
  \begin{tabular}{l c c c}
\hline
Source of uncertainty &  INEL & NSD-enhanced &
    SD-enhanced\\\hline
    HF energy scale & 10\% & 10\%  & 10\% \\
    \Cas energy scale & 17\%  & 17\% &  17\% \\
    Noise and \PU & $\approx$$10^{-3}$ & $\approx$$10^{-3}$ & $\approx$$10^{-3}$ \\
    Event selection & 0.7\% & 0.01\% & 5\% \\
    Energy threshold in calorimeter towers & 1\% & 1\% & 1\% \\
    Model dependence & $<$3.5\% & $<$3.5\% & $16-37\%$\\
    Statistical & $<$1\% &  $<$1\% &  $<$1\% \\
    \hline
  \end{tabular}
\label{sysunc}
\end{table*}

\subsection{Collision counting, noise, and \PU}

The number of selected events in the analysis, $N_\mathrm{sel}$, is
corrected to eliminate  the residual
contribution from  detector noise to yield the corrected number of
events, $N_\text{corr}$, containing only signal and no noise events.
In the following a fundamental and comprehensive discussion of event
counting is provided despite the fact that the final corrections are
just on the percent level.
With $N_\text{ZB}$ and $N_\text{EB}$ being the number of events collected
with the unbiased and empty-beam triggers, respectively, and
$f_\text{ZB}$ and $f_\text{EB}$ the corresponding fractions of
offline-selected events, we can define the number of selected
collision events $N_\text{sel}=N_\text{ZB}f_\text{ZB}$, and the
number of noise events in the same data sample
$N_\text{noise}=N_\text{ZB}f_\text{EB}$. The  latter contains
$N_\text{sig+noise}=N_\mathrm{corr}f_\mathrm{EB}$ events that are
selected because towers in the same event are above threshold due to
signal and noise fluctuations. Thus, the corrected number of events
containing collisions is
\begin{equation}
\begin{split}
N_\text{corr} &= N_\text{sel} - N_\text{noise} + N_\text{sig+noise} \\ &=  \frac{N_\text{ZB}(f_\text{ZB} -  f_\text{EB})}{1-f_\text{EB}}
\\ &= N_\text{ZB} f_\text{ZB} p,
\end{split}
\end{equation}
where we define the purity as $p=(1-f_\text{EB}/f_\text{ZB})/(1-f_\text{EB})$.
The purity of the data used in this analysis is found to be above
99\%. The noise contribution depends weakly on the
event selection criteria.

The reconstructed number of collisions is also corrected for the effect
of \PU. The number of proton-proton interactions per bunch crossing $n$ follows
a Poisson distribution with a
mean value $\lambda\epsilon$, where $\epsilon$ is
the probability for each collision to be observed. The probability to have no interaction is
given by $\text{e}^{-\lambda\epsilon}=1-N_\mathrm{corr}/N_\mathrm{ZB}$, which allows $\lambda$ to be
determined from inelastic events in data.
Here we find $\lambda=-\ln(1-f_\text{ZB}p)/\epsilon=0.055\pm0.001$, using the
value of $f_\text{ZB}$
determined from the INEL event selection, and $\epsilon$ from simulations
(see also Table~\ref{tab:corr}). The uncertainty is driven by the model dependence
of $\epsilon$ of about 2\%.

The number of visible collisions in $N_\text{tot}$ bunch crossings
is $N_\text{vis}=N_\text{tot}\sum_{n=0}^\infty
n\pois{n}{\lambda\epsilon}=N_\text{tot}\lambda\epsilon$.
In the presence of \PU another important quantity is the probability for the
observation of events with exactly $n$ simultaneous collisions, $\epsilon_n=1-(1-\epsilon)^n$.
The number of actually observed events is then
$N_\text{obs}=N_\text{tot}\sum_{n=0}^\infty \epsilon_n \pois{n}{\lambda}$.
Using this result we can correct for \PU using the factor
\begin{equation}
f_\text{PU} = \frac{N_\text{vis}}{N_\text{obs}}=\epsilon\lambda \left(\sum\limits_{n=0}^\infty\epsilon_n\pois{n}{\lambda}\right)^{-1}=\frac{\epsilon\lambda}{1-\re^{-\epsilon\lambda}}.
\end{equation}
For the data analysis we use the corrected number of collisions
\begin{equation}
\label{eq:ncoll}
N_\text{coll}=N_\mathrm{ZB}f_{ZB}pf_\mathrm{PU}=-N_\mathrm{ZB}\ln\frac{1-f_\mathrm{ZB}}{1-f_\mathrm{EB}}
\end{equation}
for Eq.~(\ref{eq:enfl}).  The same expression can also be obtained
by arguing that during no-beam data taking the average number of
collisions per event is $\lambda_\mathrm{EB}=-\ln(1-f_\mathrm{EB})$
whereas during normal data taking it is
$\lambda_\text{coll}+\lambda_\mathrm{EB}=-\ln(1-f_\mathrm{ZB})$. After
inserting into
$N_\text{coll}=N_\mathrm{ZB}\lambda_\text{coll}$ this is
identical to Eq.~(\ref{eq:ncoll}).
In the final expression only
$f_\mathrm{EB}$ and $f_\mathrm{ZB}$ are relevant, thus, the parameters
$p$ and $f_\mathrm{PU}$ are intermediate quantities highlighting the individual
importance of noise and pileup corrections.
It must also be highlighted that the
efficiency $\epsilon$ does not enter the final result.

In general, the impact of \PU depends
on the event selection procedure. In particular, an exclusivity
criterion as used in the SD-enhanced category leads to fewer
selected events in the presence of a larger number of simultaneous
collisions.  Using the corrected number of inelastic collisions,
$N_\text{INEL}$, and the corrected number of collisions inclusively selected by the HF$+$,
$N_{\text{HF}+}$, or by the HF$-$, $N_{\text{HF}-}$, the number of SD-enhanced collisions is calculated
from $N_\text{SD}=2N_\text{INEL}-N_{\text{HF}-}-N_{\text{HF}+}$. For
NSD-enhanced collisions this relation is
$N_\text{NSD}=N_{\text{HF}-}+N_{\text{HF}+}-N_\text{INEL}$.
The results from this collision counting procedure are summarised in Table~\ref{tab:corr}.
The combined corrections for
each category are at the level of 1\%. The value quoted for
$\epsilon$ is the average obtained from the different event generators
with a maximum discrepancy between the model predictions of about 2\%.
The maximum uncertainty of deriving
$p\,f_\mathrm{PU}$ is less than $<10^{-3}$.

\subsection{Energy measurement}

The measured response from the calorimeters is corrected to
the stable-particle level
  to provide a well-defined event classification and energy quantification
for comparisons to the model predictions.
The corrections are applied explicitly for each
range in pseudorapidity. There is no relevant migration or detector smearing in
pseudorapidity; it is basically the characteristic response of the calorimeters
as well as the event selection acceptance and inefficiency that are corrected.
These corrections are determined with the \PYTHIA{8} tune
\monash, \PYTHIA{8} tune \textsc{4c} with \textsc{mbr} model,
\EPOS, and \QGSJET simulated event samples.
 The corrections are evaluated from the ratio of the predictions at
 the stable-particle level
  to the predictions at the detector level for every $\abs{\eta}$ bin. The final correction is the average of the
four different simulated samples. The magnitude of the correction varies from $1.5$ to around $2.5$
depending on the value of $\abs{\eta}$  and the selection criteria applied at the stable-particle
level. The main contribution to the correction is related to the
extrapolation of observed detector-level energy above the calorimeter
noise threshold to the energy with no threshold applied at the stable-particle level.

\section{Uncertainties}

The energy scales for the HF and \Cas calorimeters are known to within
an accuracy of $\errscalehf\%$~\cite{Chatrchyan:2011wm} and
$\errscalecas\%$~\cite{Andreev:2010zzb}, respectively. These are the
dominant sources of experimental uncertainty in this analysis.

The impact of the energy scale uncertainty on the
measurement of the energy density is estimated
by scaling the tower energies up and down by the energy scale uncertainties
in the data while keeping the simulated correction factors
constant. The resulting impact is $\errscalehf\%$ for HF and
$\errscalecas\%$ for \Cas as expected.

{\tolerance=2000
To assess the residual impact of detector noise on the event selection,
the thresholds in the event selection at detector level
are increased from
5 to 5.5\GeV for all INEL, NSD-enhanced, and SD-enhanced categories.
This corresponds to
an improved noise rejection at the expense of larger correction factors.
The resulting
uncertainties are about 0.7, 0.01, and 5\% for the INEL, NSD-enhanced,
and SD-enhanced categories, respectively.
}

Furthermore, to study the impact of the
energy threshold on the energy measurement, the threshold for the tower
energy sum is increased by the energy scale uncertainty,
which leads to uncertainties of 1\% for all three categories.

{\tolerance=800
The systematic uncertainty due to model dependence is estimated from the
maximum variation of the correction factor values obtained using
the event generators \PYTHIA{8}
with \textsc{monash} and \textsc{4c+mbr} tunes, \EPOS, and \QGSJET.
The resulting
uncertainty is below 3.5\% for the INEL and
NSD-enhanced categories, while
for the SD-enhanced category it varies from 16 to 37\%, depending on $\eta$.
\par}

The statistical uncertainty is $<1$\%, which is
significantly smaller than the systematic uncertainties.

The individual contributions for each $\abs{\eta}$ bin are assumed to
contribute quadratically to the total systematic uncertainty since the
contributions are
not correlated within a bin; the systematic uncertainties are, however,
highly correlated between different $\abs{\eta}$ bins. All uncertainties
are summarised in Table~\ref{sysunc}.

\section{Results}

\begin{figure*}[pb]
  \centering
    \includegraphics[width=0.49\textwidth]{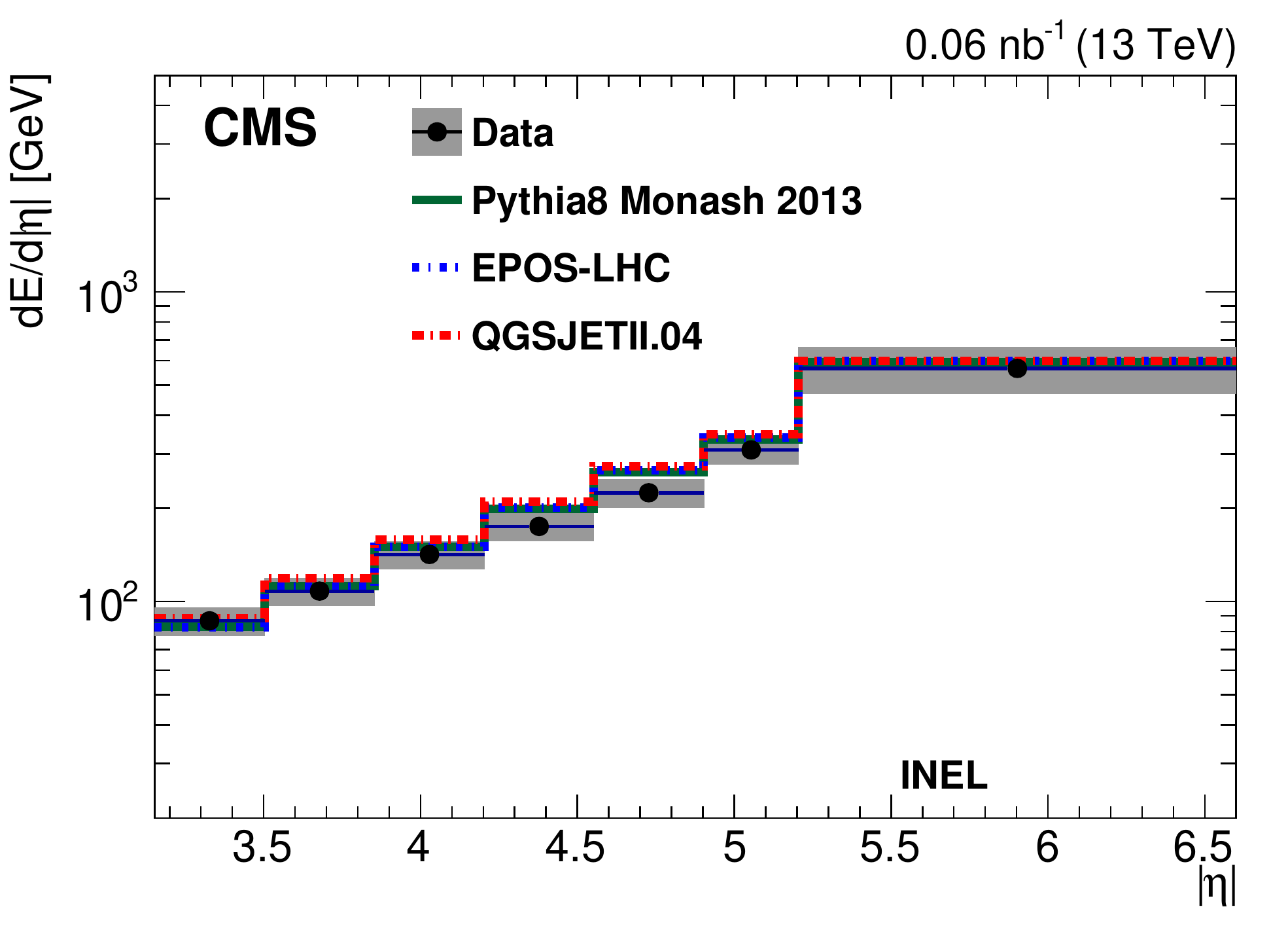}
    \includegraphics[width=0.49\textwidth]{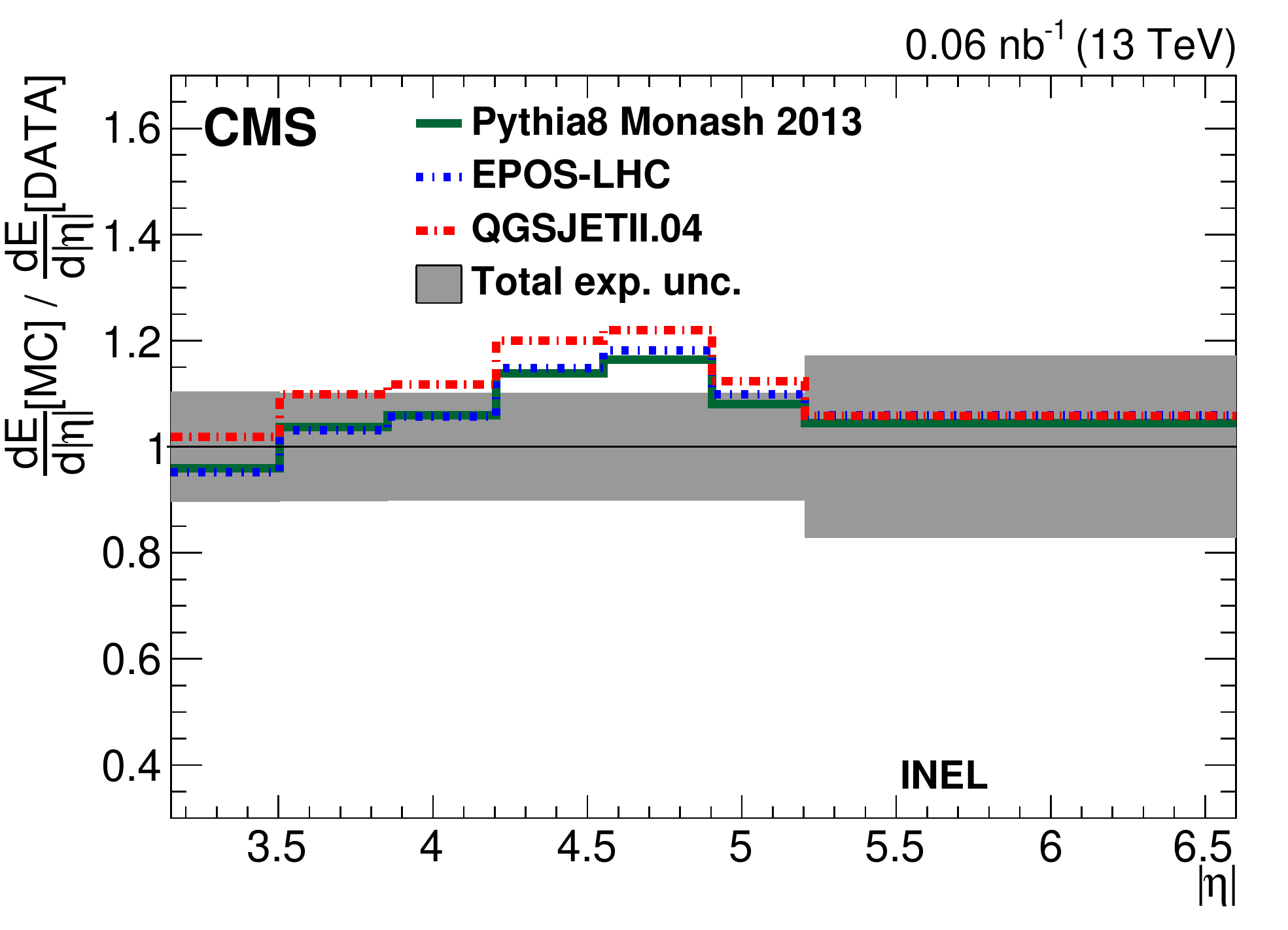}
  \includegraphics[width=0.49\textwidth]{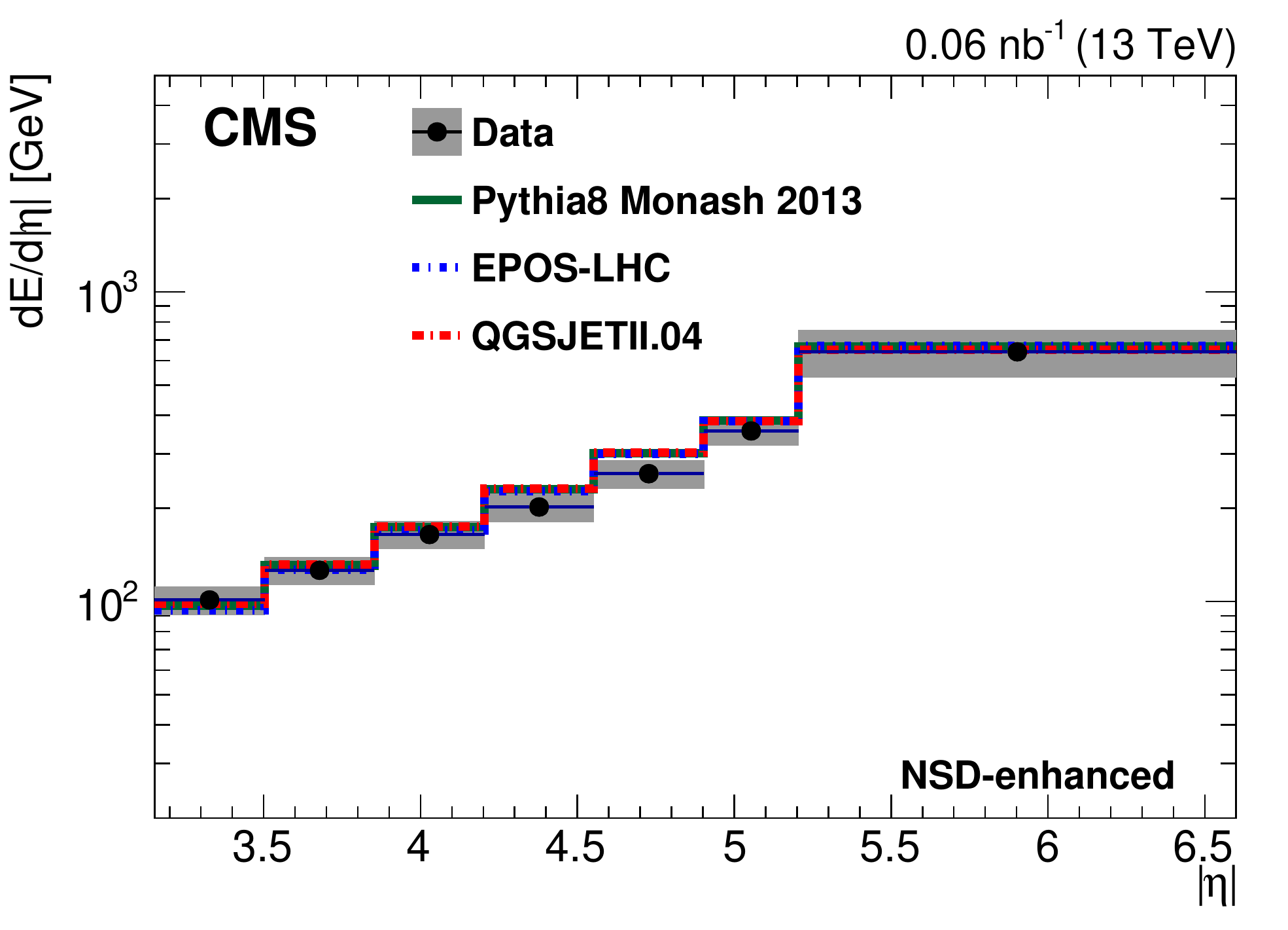}
  \includegraphics[width=0.49\textwidth]{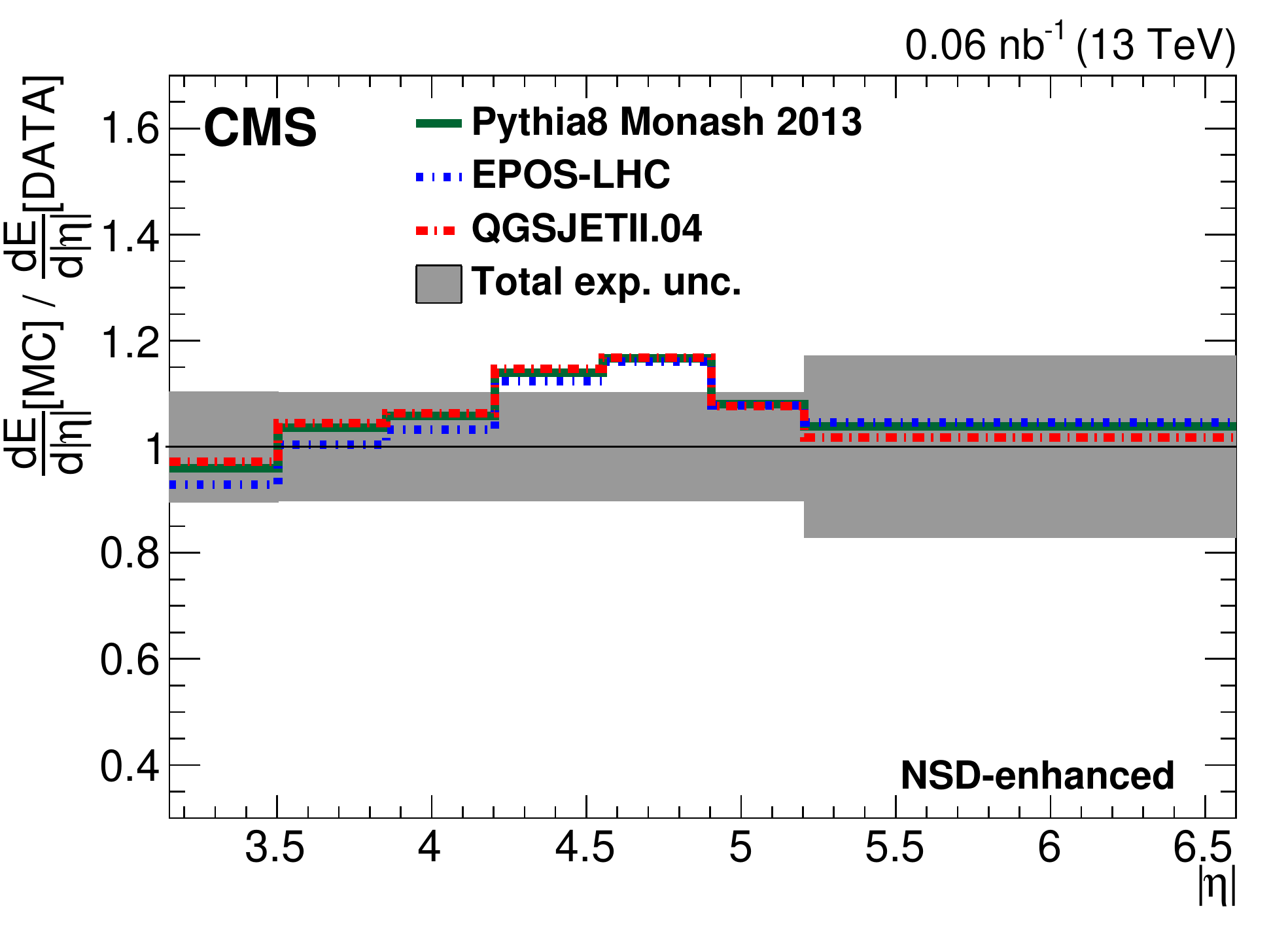}
\includegraphics[width=0.49\textwidth]{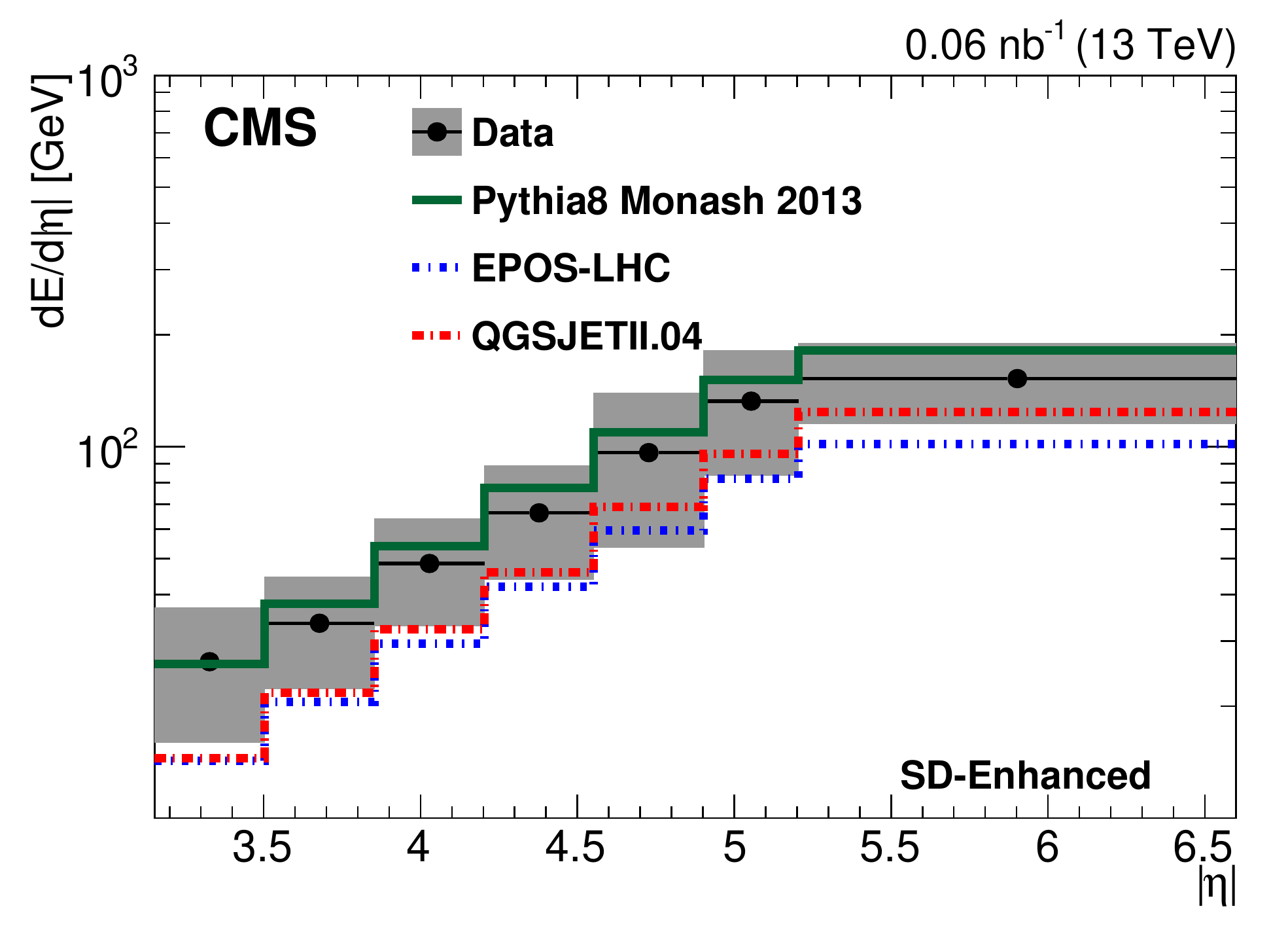}
  \includegraphics[width=0.49\textwidth]{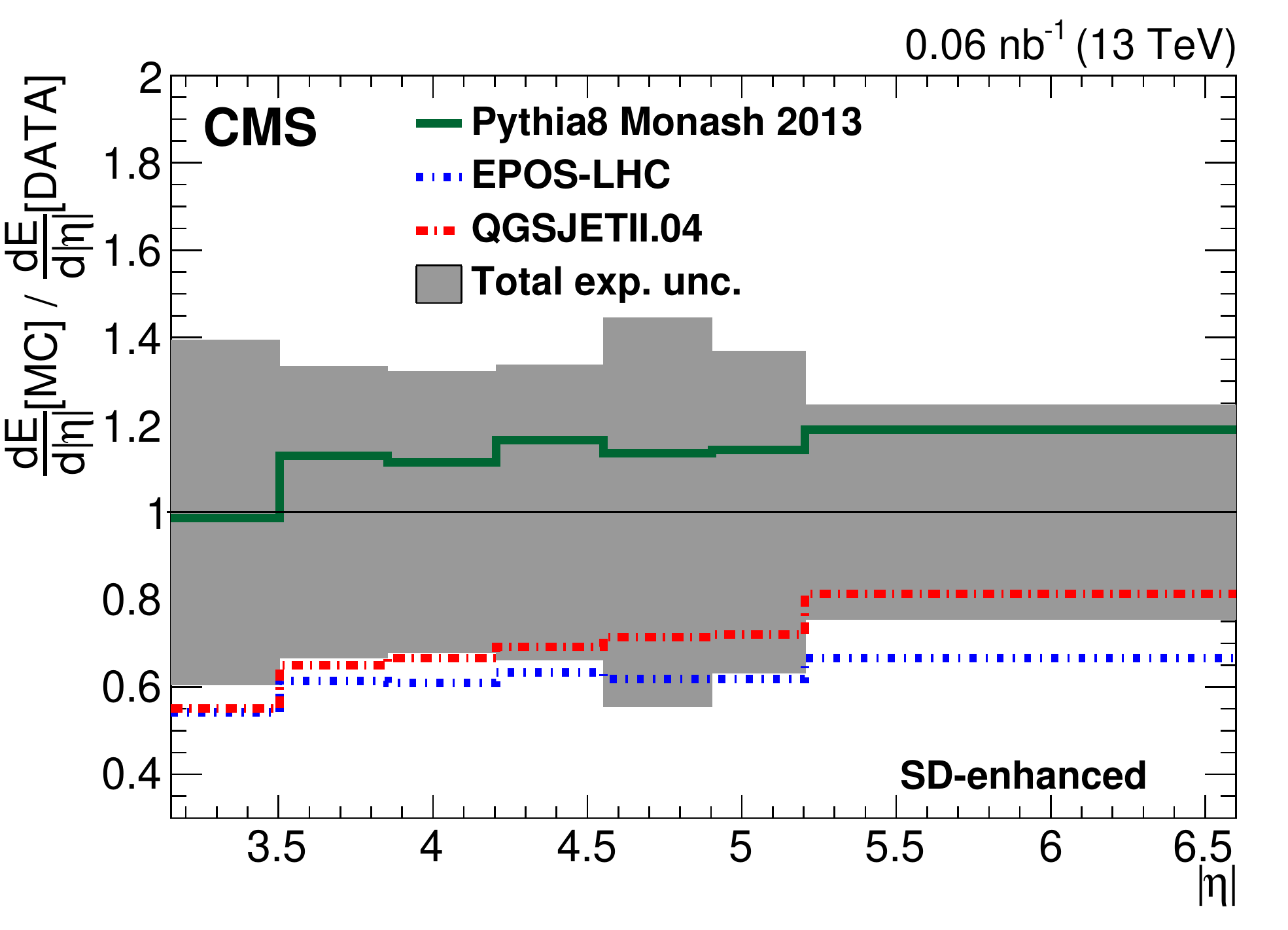}
  \caption{Energy density at the stable-particle level
   for the INEL (upper row), NSD-enhanced (middle row), and SD-enhanced (lower row) categories
    compared
    to predictions from \PYTHIA{8} \textsc{monash}, \EPOS, and \QGSJET.
     The gray band shows the total systematic
    uncertainty. The right panels show the ratio of
     model predictions to measured data.\label{figii}}
\end{figure*}

\begin{figure*}[ptb]
  \centering
     \includegraphics[width=0.49\textwidth]{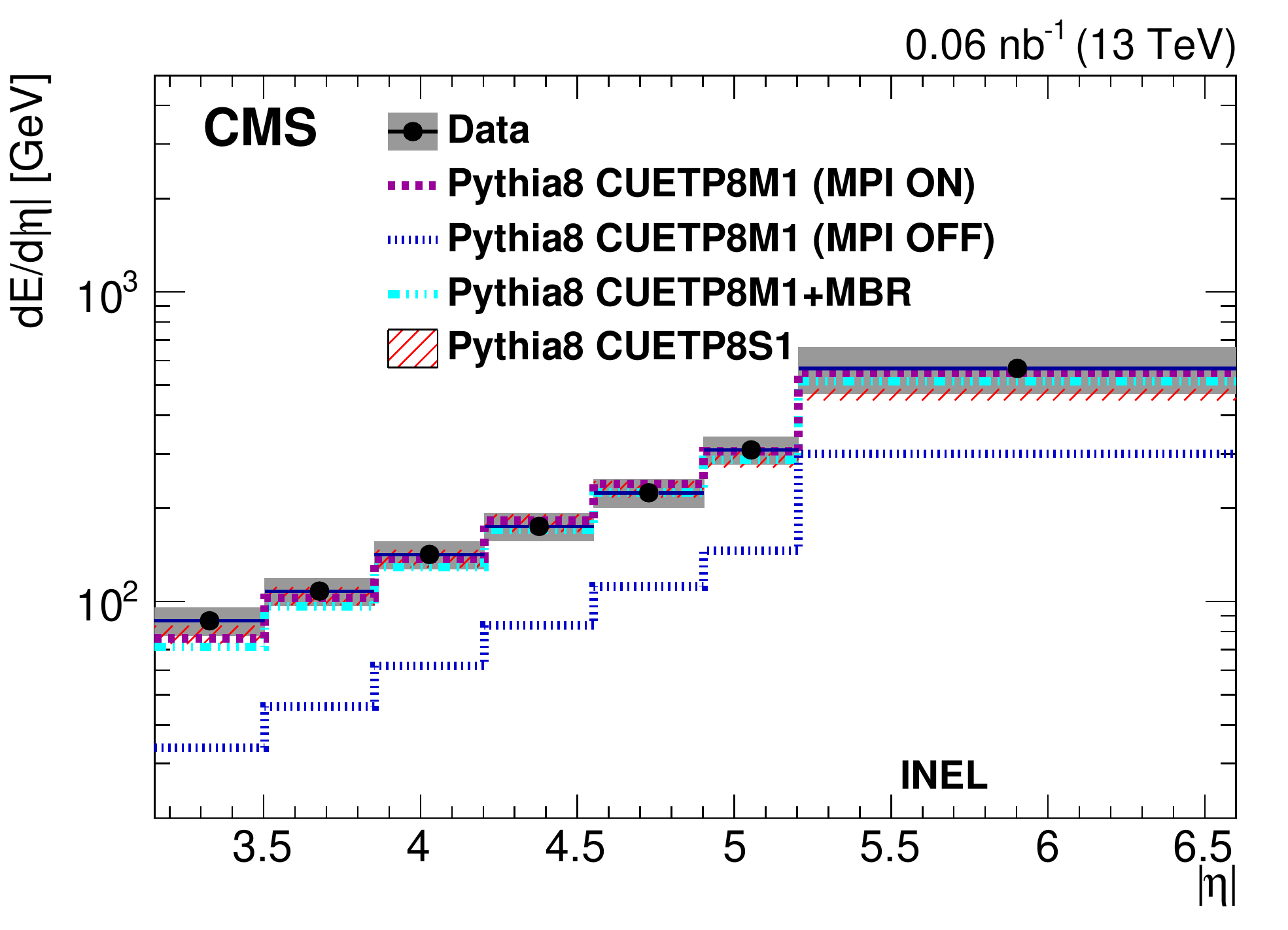}
    \includegraphics[width=0.49\textwidth]{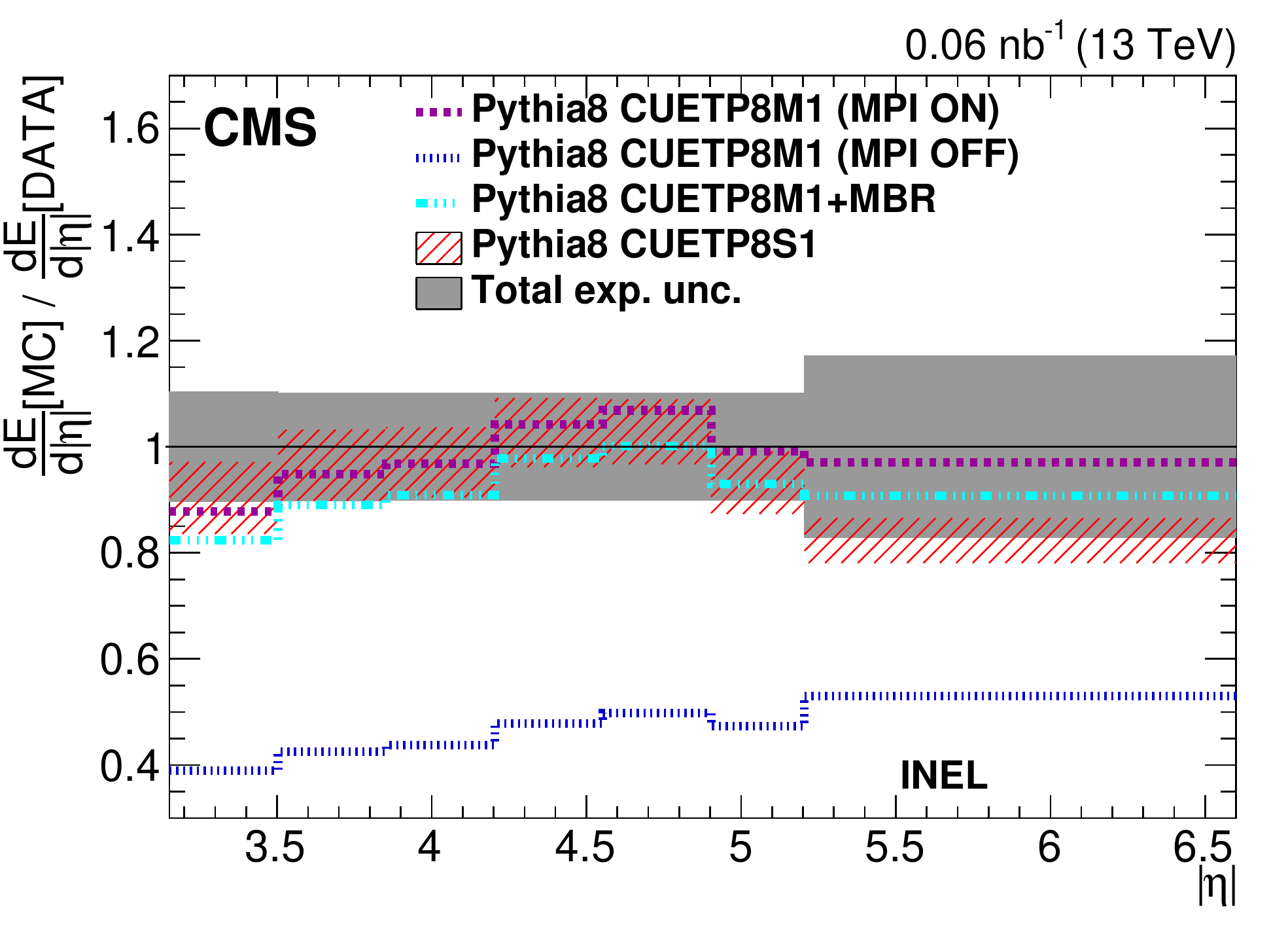}
\includegraphics[width=0.49\textwidth]{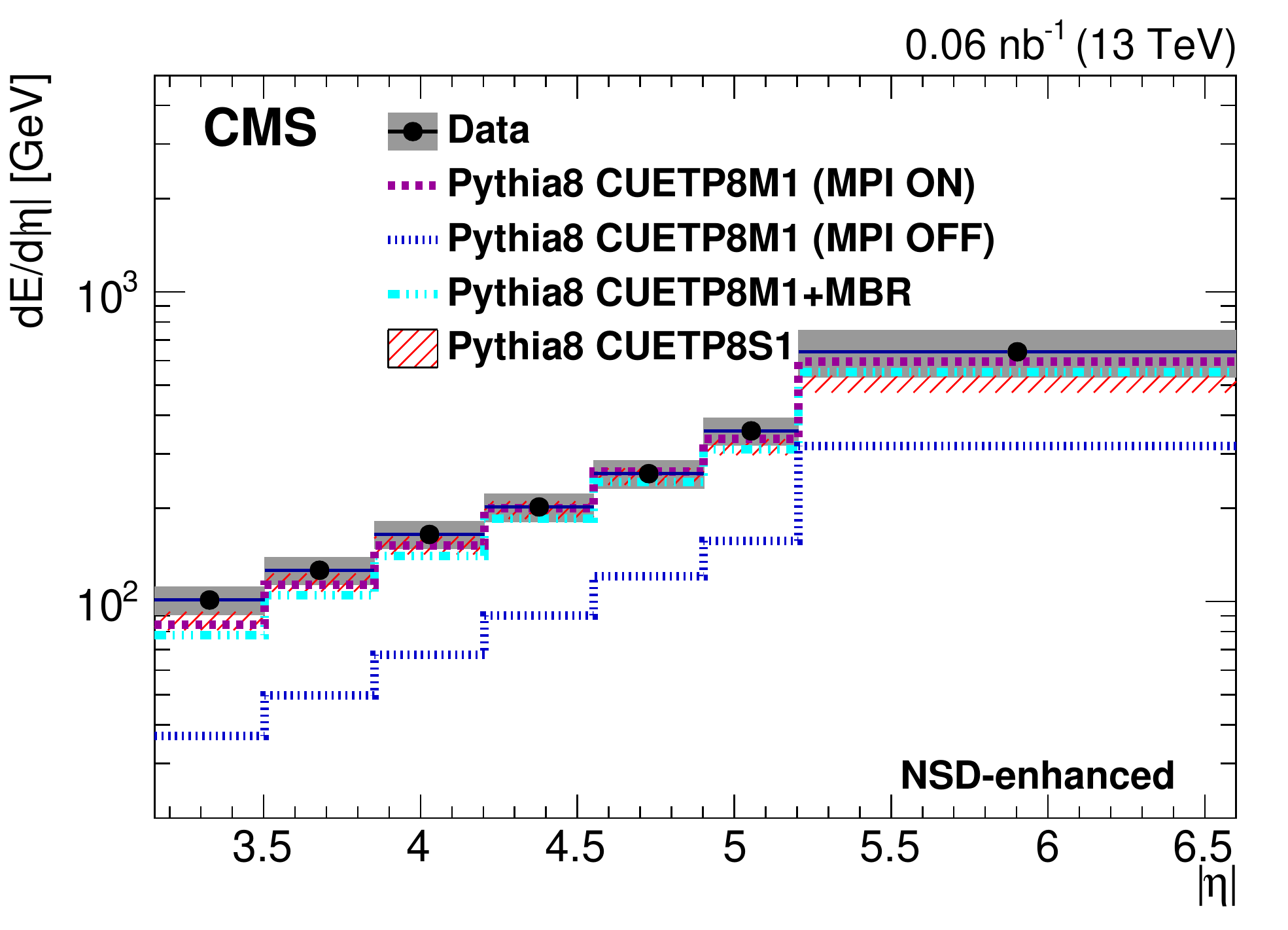}
  \includegraphics[width=0.49\textwidth]{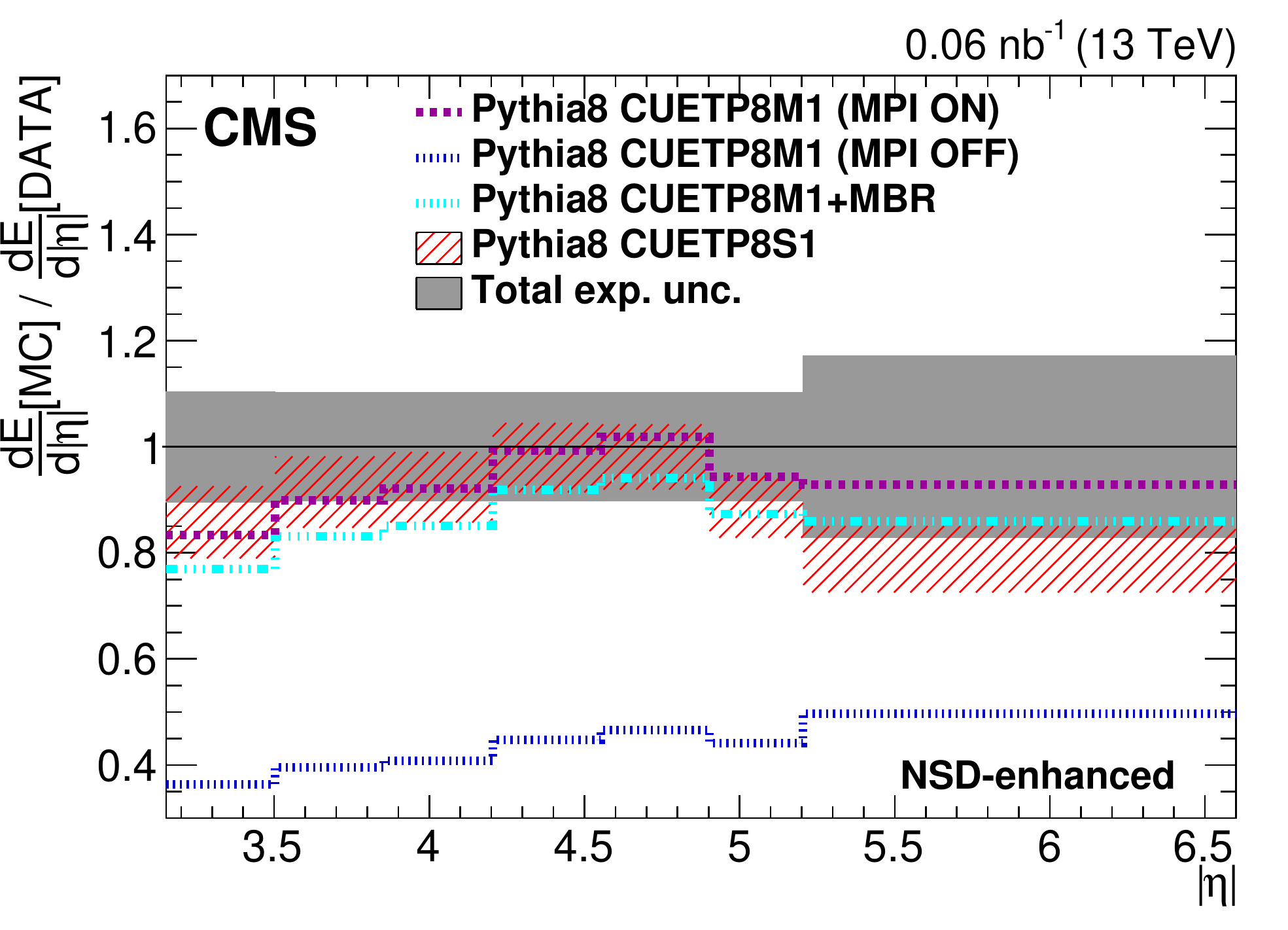}
\includegraphics[width=0.49\textwidth]{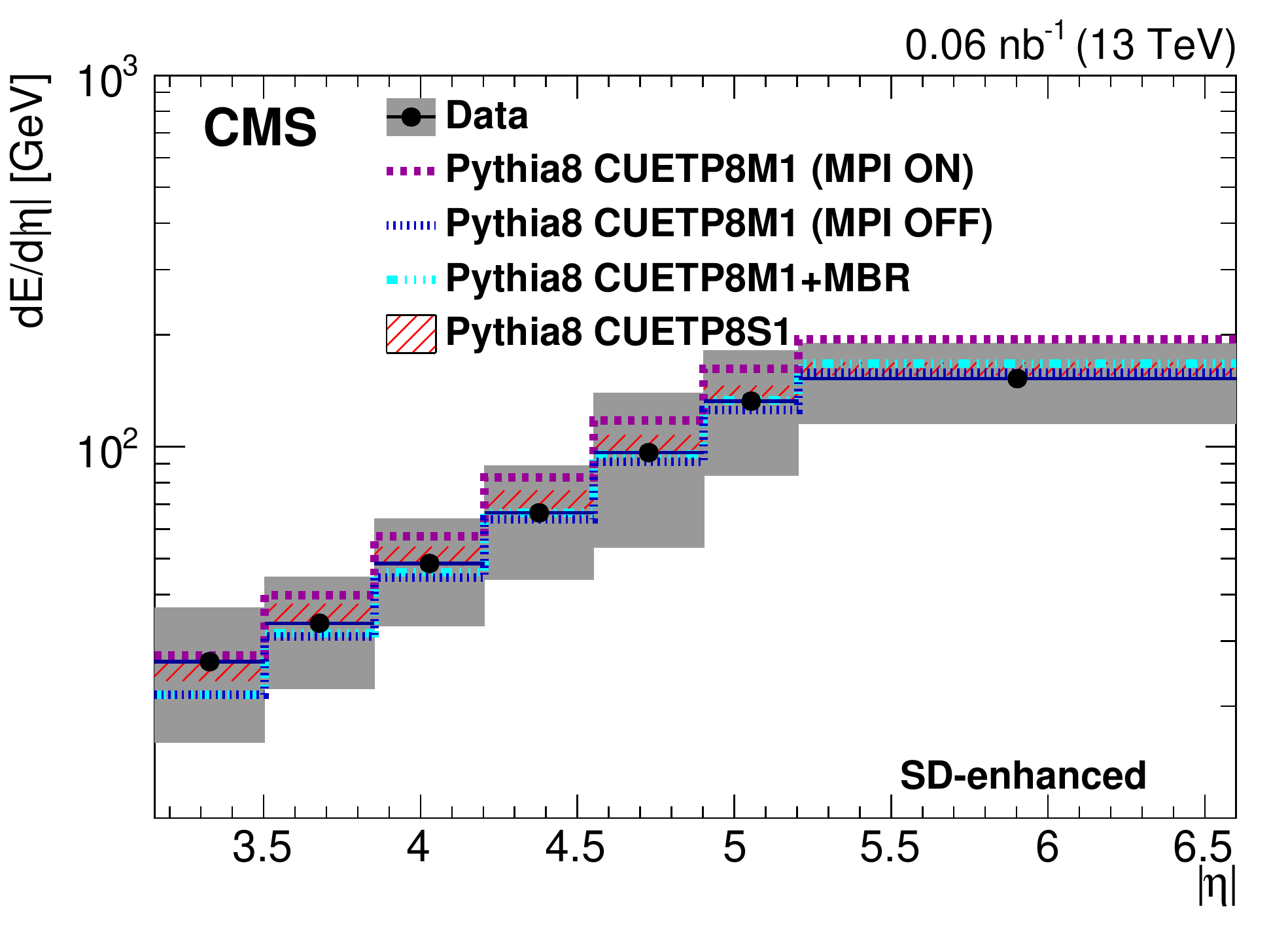}
  \includegraphics[width=0.49\textwidth]{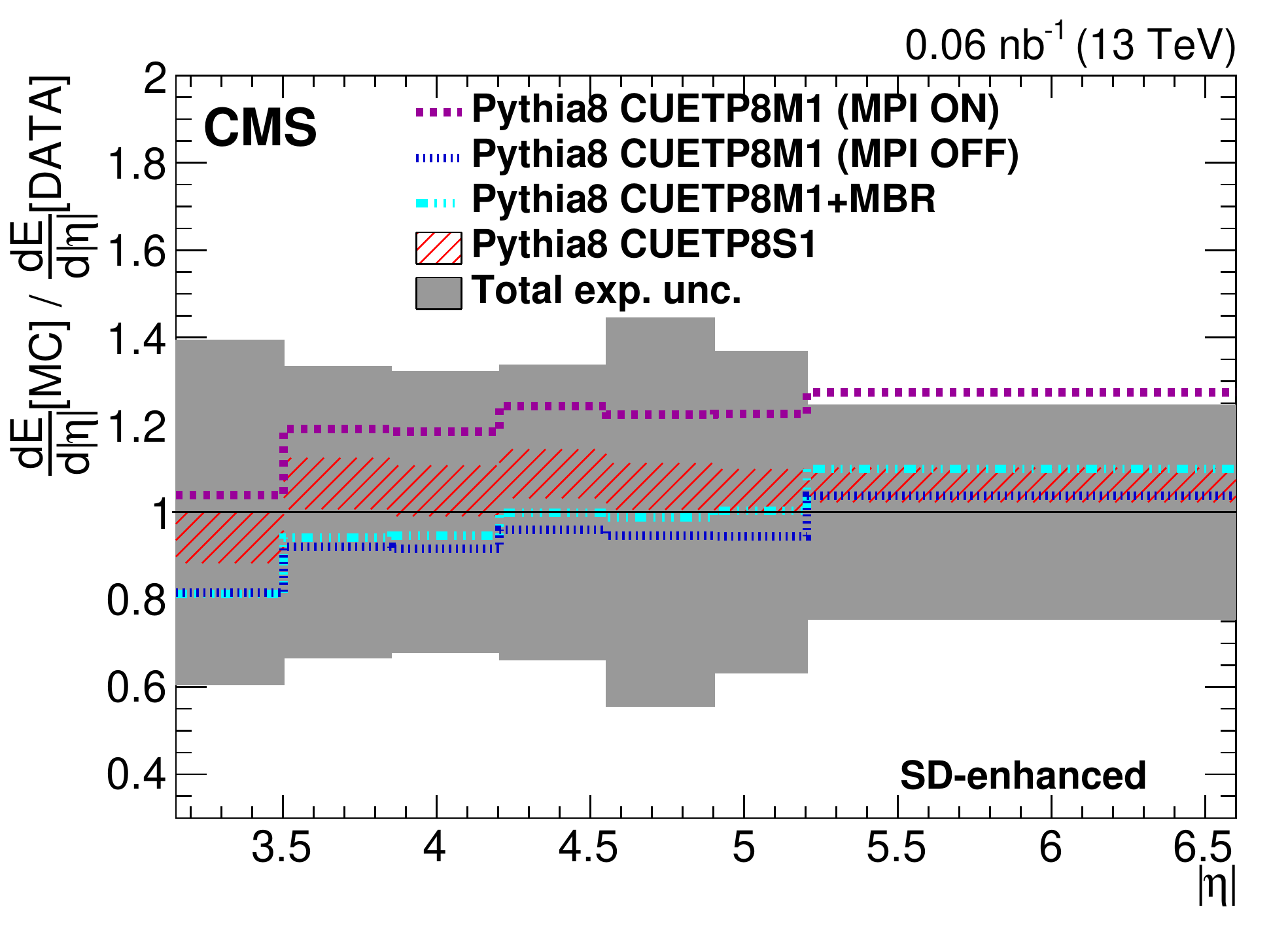}
  \caption{Energy density at the stable-particle
   level for the INEL (upper row), NSD-enhanced (middle row), and SD-enhanced (lower row) categories
  compared
    to predictions from \PYTHIA{8} with the tunes \textsc{cuetp8m1},
    \textsc{cuetp8m1+mbr}, and \textsc{cuetp8s1}.
       The gray band shows the total systematic
    uncertainty.
    The band around \PYTHIA{8} \textsc{cuetp8s1} corresponds to
    the uncertainties of the tune
    parameters.  The right panels show the ratio of
     model predictions to measured data.\label{figiii}}
\end{figure*}

\begin{figure}[tb]
 \centering
  \includegraphics[width=\linewidth]{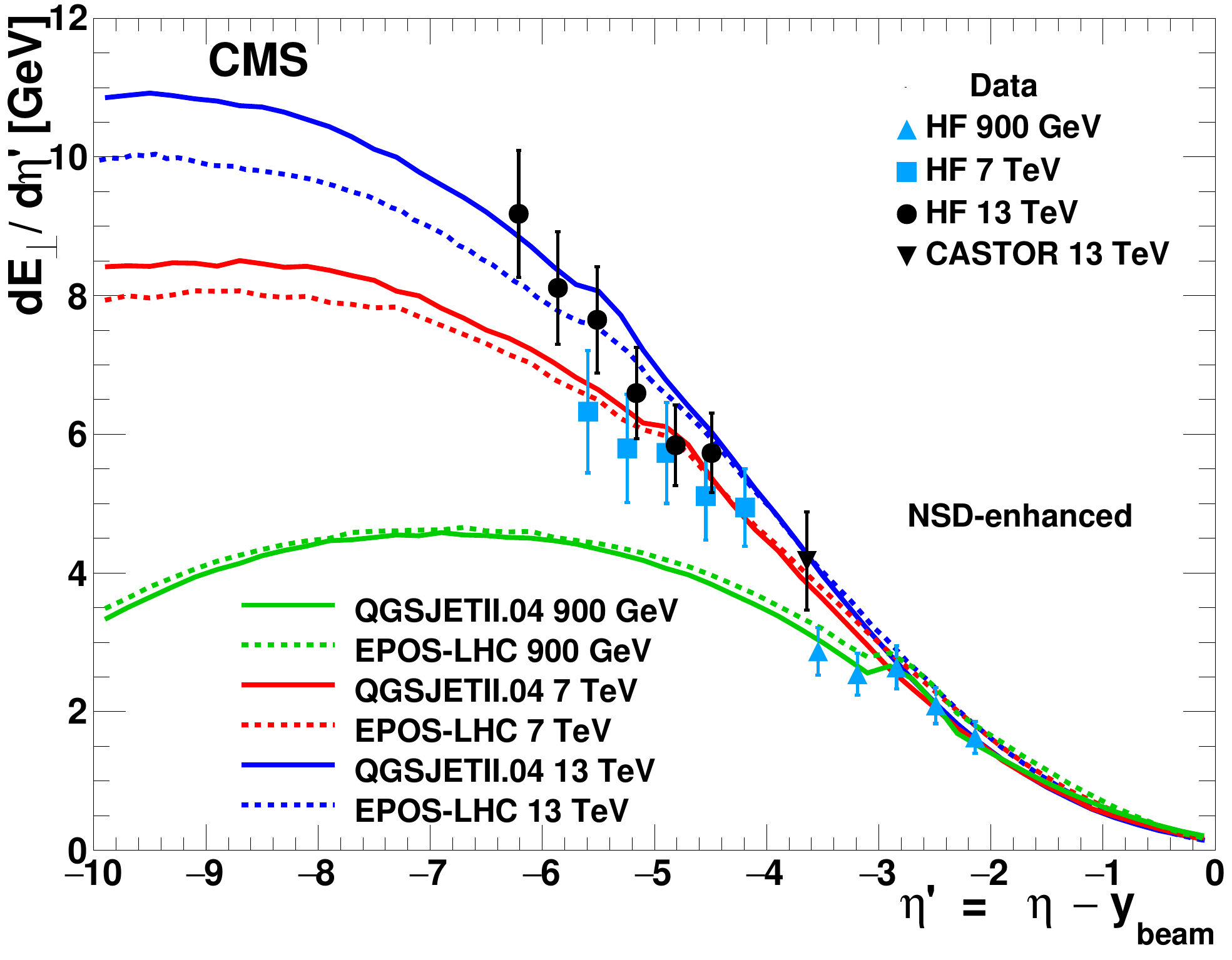}
  \caption{
 A comparison of the measurements of the transverse energy density,
 $\dhi{\Et}{\eta'}$, at $\sqrts=13\TeV$, as a function of shifted
 pseudorapidity,
 $\eta'=\eta-\yb$, to the predictions and to earlier
 proton-proton data~\cite{Chatrchyan:2011wm} for an
 NSD-enhanced selected sample
 at several different centre-of-mass energies.
 The error bars indicate the
 total systematic uncertainties. The beam rapidities $\yb$ are
 about 9.5, 8.9, and 6.8
 at $\sqrts$ of 13, 7 and 0.9\TeV, respectively.}
 \label{lk05}
\end{figure}

The measured energy density, $\dhi{E}{\eta}$, in the range
$-6.6<\eta<-5.2$ and $3.15<\abs{\eta}<5.20$, corrected to the
stable-particle level,
is presented in Figs.~\ref{figii} and \ref{figiii}.

{\tolerance=800
A comparison of the measured average energy density to model predictions for
the INEL selection is shown in
Figs.~\ref{figii}\,(upper) and \ref{figiii}\,(upper).  The gray
band represents the total systematic uncertainty correlated across
$\abs{\eta}$ bins.  The statistical uncertainties are $<$1\% and
are not shown. In the left panel the comparison of the distribution in data and simulation is shown,
while in the right panel the ratio quantifies the agreement between them.
While the cosmic
ray models (\EPOS and \QGSJET) and the \PYTHIA{8} \textsc{monash} tune describe the data
well at $\abs{\eta}<4$ and in the \Cas region, they overshoot the data
around $\abs{\eta}\approx4.5$. This is most pronounced in
\QGSJET. The \PYTHIA{8} \textsc{cuet} tunes describe the data slightly better,
but have a tendency to
undershoot the data towards $\abs{\eta}<3.5$.
The band around \PYTHIA{8} \textsc{cuetp8s1} in
Fig.~\ref{figiii} indicates the typical
uncertainties due to the tune parameters. The best description of the
data is provided by the \PYTHIA{8} tune \textsc{cuetp8s1}.
When MPIs are switched off in \PYTHIA{8} more than half of the
measured energy is missing, with a slight dependence on $\eta$.
}

In Figs.~\ref{figii}\,(middle) and \ref{figiii}\,(middle) the energy
density measurements are compared with predictions for the NSD-enhanced category.
The differences between the model predictions are
smaller compared with the INEL category. The \EPOS and \QGSJET hadronic
event generators overshoot the measurement only at $\abs{\eta}\approx4.5$ and otherwise
show a good description of the data.
The \PYTHIA{8} tune \textsc{cuetp8s1} at the upper limit
of its uncertainties provides the best overall description of the data.

{\tolerance 800
Figure~\ref{figii}\,(lower) shows a comparison of the
energy density measurements
as a function of $\eta$ for the SD-enhanced category to
predictions from \PYTHIA{8} \textsc{monash}, \EPOS, and \QGSJET.
The comparison of the same data to the different \PYTHIA{8}
tunes is shown in Fig.~\ref{figiii}\,(lower).
For the SD-enhanced category the model spread becomes significantly larger. It is
interesting that the \EPOS and \QGSJET models are both compatible with the data
only at the very lower limit of the systematic uncertainties,
while all \PYTHIA{8} tunes are consistent with
the data within the uncertainties. Furthermore, the shape of all the model
predictions is very similar and, in contrast to the INEL and NSD-enhanced data,
consistent with the data. Finally, we observe that for the SD-enhanced
category switching off MPIs
in simulations has almost no impact on the model predictions.
This is an indication that the influence of MPIs
within the diffractive system is
small, whereas MPIs between the colliding protons will quickly destroy the \SD
signature. Thus, the SD-enhanced event selection is an effective way
to minimise MPI effects.
\par}

For a detailed comparison to previously published energy density
results at lower centre-of-mass energies~\cite{Chatrchyan:2011wm}, the event selection
is adapted to match the one previously used at detector and stable-particle
levels.
The
whole measurement is repeated for the NSD-enhanced category with
the requirement of at least one charged particle on both sides of the
interaction point in the pseudorapidity range $3.9<\abs{\eta}<4.4$.
This is combined with a reduced energy threshold of $4\GeV$  to ensure
consistency. Finally, for all calculations the transverse
energy $\Et=E \cosh(\eta)$ per tower is used instead of just the tower
energy $E$. In Fig.~\ref{lk05} the resulting corrected transverse energy density, $\dhi{\Et}{\eta'}$, is
compared to earlier published CMS data at lower $\sqrts$ and
to model predictions, as a function of the shifted pseudorapidity variable
$\eta'=\eta-\yb$.
The analysis presented here uses the latest CMS detector
description in the simulations, which includes an improved knowledge of
the HF nonuniformity due to nonsensitive
areas~\cite{Wohrmann:2013nta}, that was not present in the original publication~\cite{Chatrchyan:2011wm}.
In order to facilitate
the direct comparison of the current analysis with earlier results~\cite{Chatrchyan:2011wm}, corrections
are applied to the published data that cause the results in the HF to be
shifted in an $\eta$-dependent way; from about $-2\%$ at $\abs{\eta}=3$ to
about $-15\%$ at $\abs{\eta}=5$, which is within the experimental uncertainties of these data.

A comparison of the model predictions and data at different $\sqrts$
is shown in Fig.~\ref{lk05}. Both the data and the model predictions are shifted
by the beam rapidity to $\eta'=\eta-\yb$.
The data are consistent with longitudinal scaling
within the experimental uncertainties.
The observed behaviour is in agreement with the measurements of
earlier experiments in proton-proton and heavy ion collisions (\eg~\cite{Wohrmann:2013nta}).
At $\eta'\approx0$ the transverse energy density does not depend on $\sqrts$,
which is in agreement with the hypothesis of limiting fragmentation.

\section{Summary}

{\tolerance 800
The energy density, $\dhi{E}{\eta}$, is measured in the pseudorapidity range
$-6.6<\eta<-5.2$ and $3.15<\abs{\eta}<5.20$. Special low-luminosity data
recorded by the CMS experiment during proton-proton collisions at
the centre-of-mass energy $\sqrts=13\TeV$ are analysed for this purpose. The data are presented at the stable-particle
  level to allow a straightforward comparison to any theory
prediction or model simulation. The measurements are compared to models tuned to describe high-energy
hadronic interactions (\PYTHIA{8}) and to the predictions
of models used in cosmic ray physics (\EPOS, \QGSJET)
for inclusive inelastic (INEL), \NSD (NSD-enhanced), and \SD (SD-enhanced) event selection categories.

It is shown that the INEL and NSD-enhanced categories are extremely sensitive to
multi-parton interactions, while the SD-enhanced category is
essentially unaffected. The shape of the
measured $\eta$ dependencies suggest a difference in the
models compared to the data. However, the predictions of \PYTHIA{8}
tune \textsc{cuetp8s1} are in satisfactory agreement with all
measurements when the experimental and tune uncertainties are
combined. The \EPOS and \QGSJET models exhibit the largest differences when compared to the \SD results.
\par}

At high energies, the hypothesis of limiting
fragmentation~\cite{Benecke:1969sh,Ruan:2010ig} assumes a longitudinal
scaling behaviour in terms of shifted pseudorapidity
$\eta'=\eta-\yb$ (where $\yb$ is the beam rapidity)
and thus soft-particle production in the
projectile fragmentation region, $\eta'\approx0$, is predicted to be
independent of the centre-of-mass energy. This is studied by
measuring the transverse energy density $\dhi{\Et}{\eta}$, with
$\Et=E \cosh(\eta)$, and comparing it to measurements performed in
proton-proton collisions at different centre-of-mass energies.
The predictions of
the \EPOS and \QGSJET models nicely describe the combined data in the
forward pseudorapidity range close to the projectile fragmentation
region. The result supports the mechanism of
limiting fragmentation. Since this
predicts the independence of very forward particle production on the
energy of the projectile particle, these data are very important for
the modelling of ultra-high energy interactions that typically occur in
cosmic ray collisions.

\begin{acknowledgments}
We congratulate our colleagues in the CERN accelerator departments for the excellent performance of the LHC and thank the technical and administrative staffs at CERN and at other CMS institutes for their contributions to the success of the CMS effort. In addition, we gratefully acknowledge the computing centres and personnel of the Worldwide LHC Computing Grid for delivering so effectively the computing infrastructure essential to our analyses. Finally, we acknowledge the enduring support for the construction and operation of the LHC and the CMS detector provided by the following funding agencies: BMBWF and FWF (Austria); FNRS and FWO (Belgium); CNPq, CAPES, FAPERJ, FAPERGS, and FAPESP (Brazil); MES (Bulgaria); CERN; CAS, MoST, and NSFC (China); COLCIENCIAS (Colombia); MSES and CSF (Croatia); RPF (Cyprus); SENESCYT (Ecuador); MoER, ERC IUT, and ERDF (Estonia); Academy of Finland, MEC, and HIP (Finland); CEA and CNRS/IN2P3 (France); BMBF, DFG, and HGF (Germany); GSRT (Greece); NKFIA (Hungary); DAE and DST (India); IPM (Iran); SFI (Ireland); INFN (Italy); MSIP and NRF (Republic of Korea); MES (Latvia); LAS (Lithuania); MOE and UM (Malaysia); BUAP, CINVESTAV, CONACYT, LNS, SEP, and UASLP-FAI (Mexico); MOS (Montenegro); MBIE (New Zealand); PAEC (Pakistan); MSHE and NSC (Poland); FCT (Portugal); JINR (Dubna); MON, RosAtom, RAS, RFBR, and NRC KI (Russia); MESTD (Serbia); SEIDI, CPAN, PCTI, and FEDER (Spain); MOSTR (Sri Lanka); Swiss Funding Agencies (Switzerland); MST (Taipei); ThEPCenter, IPST, STAR, and NSTDA (Thailand); TUBITAK and TAEK (Turkey); NASU and SFFR (Ukraine); STFC (United Kingdom); DOE and NSF (USA).

\hyphenation{Rachada-pisek} Individuals have received support from the Marie-Curie programme and the European Research Council and Horizon 2020 Grant, contract No. 675440 (European Union); the Leventis Foundation; the A. P. Sloan Foundation; the Alexander von Humboldt Foundation; the Belgian Federal Science Policy Office; the Fonds pour la Formation \`a la Recherche dans l'Industrie et dans l'Agriculture (FRIA-Belgium); the Agentschap voor Innovatie door Wetenschap en Technologie (IWT-Belgium); the F.R.S.-FNRS and FWO (Belgium) under the ``Excellence of Science - EOS" - be.h project n. 30820817; the Ministry of Education, Youth and Sports (MEYS) of the Czech Republic; the Lend\"ulet (``Momentum") Programme and the J\'anos Bolyai Research Scholarship of the Hungarian Academy of Sciences, the New National Excellence Program \'UNKP, the NKFIA research grants 123842, 123959, 124845, 124850 and 125105 (Hungary); the Council of Science and Industrial Research, India; the HOMING PLUS programme of the Foundation for Polish Science, cofinanced from European Union, Regional Development Fund, the Mobility Plus programme of the Ministry of Science and Higher Education, the National Science Center (Poland), contracts Harmonia 2014/14/M/ST2/00428, Opus 2014/13/B/ST2/02543, 2014/15/B/ST2/03998, and 2015/19/B/ST2/02861, Sonata-bis 2012/07/E/ST2/01406; the National Priorities Research Program by Qatar National Research Fund; the Programa Estatal de Fomento de la Investigaci{\'o}n Cient{\'i}fica y T{\'e}cnica de Excelencia Mar\'{\i}a de Maeztu, grant MDM-2015-0509 and the Programa Severo Ochoa del Principado de Asturias; the Thalis and Aristeia programmes cofinanced by EU-ESF and the Greek NSRF; the Rachadapisek Sompot Fund for Postdoctoral Fellowship, Chulalongkorn University and the Chulalongkorn Academic into Its 2nd Century Project Advancement Project (Thailand); the Welch Foundation, contract C-1845; and the Weston Havens Foundation (USA).
\end{acknowledgments}

\bibliography{auto_generated}
\cleardoublepage \appendix\section{The CMS Collaboration \label{app:collab}}\begin{sloppypar}\hyphenpenalty=5000\widowpenalty=500\clubpenalty=5000\vskip\cmsinstskip
\textbf{Yerevan Physics Institute, Yerevan, Armenia}\\*[0pt]
A.M.~Sirunyan, A.~Tumasyan
\vskip\cmsinstskip
\textbf{Institut f\"{u}r Hochenergiephysik, Wien, Austria}\\*[0pt]
W.~Adam, F.~Ambrogi, E.~Asilar, T.~Bergauer, J.~Brandstetter, M.~Dragicevic, J.~Er\"{o}, A.~Escalante~Del~Valle, M.~Flechl, R.~Fr\"{u}hwirth\cmsAuthorMark{1}, V.M.~Ghete, J.~Hrubec, M.~Jeitler\cmsAuthorMark{1}, N.~Krammer, I.~Kr\"{a}tschmer, D.~Liko, T.~Madlener, I.~Mikulec, N.~Rad, H.~Rohringer, J.~Schieck\cmsAuthorMark{1}, R.~Sch\"{o}fbeck, M.~Spanring, D.~Spitzbart, W.~Waltenberger, J.~Wittmann, C.-E.~Wulz\cmsAuthorMark{1}, M.~Zarucki
% ...
\vskip\cmsinstskip
% ...
1:  Also at Vienna University of Technology, Vienna, Austria\\
% ...
\end{sloppypar}
\end{document}